\documentclass[twocolumn]{aastex631}

\usepackage{soul}
\usepackage{comment}
\usepackage{makecell}
\usepackage{float}
\usepackage{subcaption}
\usepackage[switch,columnwise]{lineno}

\newcommand{\vdag}{$^\dagger$}

\shorttitle{\texttt{GWSkyNet} Multi-Class}
\shortauthors{Abbott, Buffaz et al.}
\graphicspath{{./}{figures/}}

\begin{document}

\title{\texttt{GWSkyNet-Multi} : A Machine Learning Multi-Class classifier for LIGO--Virgo Public Alerts}

\author[0000-0001-5002-0868]{Thomas C. Abbott}
\affil{McGill Space Institute and Department of Physics, McGill University, 3600 rue University, Montreal, Quebec, H3A 2T8, Canada}

\author[0000-0003-2205-2912]{Eitan Buffaz}
\affil{McGill Space Institute and Department of Physics, McGill University, 3600 rue University, Montreal, Quebec, H3A 2T8, Canada}

\author[0000-0001-7815-7604]{Nicholas Vieira}
\affil{McGill Space Institute and Department of Physics, McGill University, 3600 rue University, Montreal, Quebec, H3A 2T8, Canada}

\author[0000-0003-4059-4512]{Miriam Cabero}
\affiliation{Division of Physics, Mathematics, and Astronomy, University of British Columbia, Vancouver, BC V6T 1Z4, Canada}

\author[0000-0001-6803-2138]{Daryl Haggard}
\affil{McGill Space Institute and Department of Physics, McGill University, 3600 rue University, Montreal, Quebec, H3A 2T8, Canada}

\author[0000-0003-2242-0244]{Ashish Mahabal}
\affiliation{Center for Data Driven Discovery, California Institute of Technology, Pasadena, CA 91125, USA}

\author[0000-0003-0316-1355]{Jess McIver}
\affiliation{Division of Physics, Mathematics, and Astronomy, University of British Columbia, Vancouver, BC V6T 1Z4, Canada}

\begin{abstract}

Compact object mergers which produce both detectable gravitational waves and electromagnetic emission can provide valuable insights into the neutron star equation of state, the tension in the Hubble constant, and the origin of the \textit{r}-process elements. However, electromagnetic follow-up of gravitational wave sources is complicated by false positive detections, and the transient nature of the associated electromagnetic emission. \texttt{GWSkyNet-Multi} is a machine learning model that attempts facilitate EM follow-up by providing real-time predictions of the source of a gravitational wave detection. The model uses information from Open Public Alerts (OPAs) released by LIGO--Virgo within minutes of a gravitational wave detection. \texttt{GWSkyNet} was introduced in \cite{GWSkynet} as a binary classifier and uses the OPA skymaps to classify sources as either astrophysical or as glitches. In this paper, we introduce \texttt{GWSkyNet-Multi}, an extension of \texttt{GWSkyNet} which further distinguishes sources as binary black hole mergers, mergers involving a neutron star, or non-astrophysical glitches. \texttt{GWSkyNet-Multi} is a sequence of three one-versus-all classifiers trained using a class-balanced and physically-motivated source mass distribution. Training on this data set, we obtain test set accuracies of 93.7\% for BBH-versus-all, 94.4\% for NS-versus-all, and 95.1\% for glitch-versus-all. We obtain an overall accuracy of 93.4\% using a hierarchical classification scheme. Furthermore, we correctly identify 36 of the 40 gravitational wave detections from the first half of LIGO--Virgo's third observing run (O3a) and present predictions for O3b sources. As gravitational wave detections increase in number and frequency, \texttt{GWSkyNet-Multi} will be a powerful tool for prioritizing successful electromagnetic follow-up. 
\end{abstract}

\keywords{\texttt{GWSkyNet}, gravitational waves, machine learning, multi-messenger astrophysics}

\section{Introduction} \label{sec:intro}
 The LIGO Scientific and Virgo Collaborations (LVC) made a monumental discovery in their second observing run (O2) with the detection of the first gravitational wave (GW) originating from a binary neutron star merger. This event, known as GW170817, is the only GW source which has been associated with electromagnetic (EM) emission (\citealt{GW170817}), marking a breakthrough for GW+EM multi-messenger astronomy. Multi-messenger astronomy has particularly exciting implications in fundamental physics --- notably, joint GW and EM detections can place independent constraints on the Hubble constant (\citealt{Hubble}), constrain the neutron star equation of state (\citealt{EOS2}), and characterize the production of $r$-process elements in compact binary coalescences (\citealt{MM_astrophysics}). However, EM follow-up of GW events is hindered by high operating costs, competitive telescope time allocations, false positive GW alerts, and the short time during which the associated kilonova is detectable.
 
\begin{figure}[!ht]
    \centering
	\includegraphics[width=0.5\textwidth]{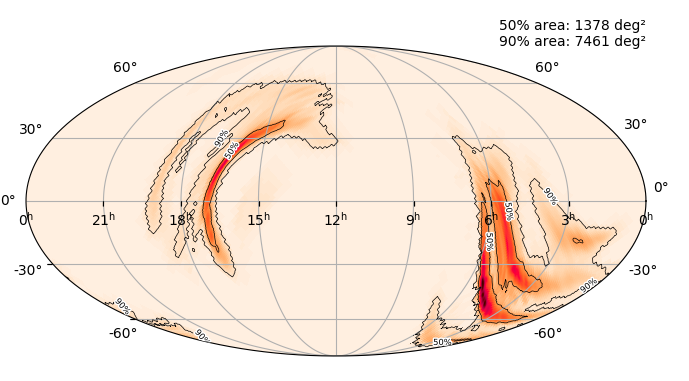}
	\caption{\textbf{Open Public Alert (OPA) skymap of S190425z}. The skymap depicts a localization region in which the source is most likely to be found, with the darker orange region describing the 50\% confidence region and the lighter orange region 90\%. The skymap images are produced by the \texttt{BAYESTAR} pipeline and made publicly available in OPAs within minutes of a gravitational wave detection. Image retrieved from GraceDB:  \url{https://gracedb.ligo.org/superevents/S190425z/view/}.}
	\label{fig: skymap}
\end{figure}

To facilitate EM follow-up, the LVC introduced the Open Public Alert (OPA) system for their third observing run (O3). OPAs are publicly available within minutes of a detection, and include a skymap (see Figure~\ref{fig: skymap}) indicating the localization region where a given event most likely originates from (\citealt{Public_Alerts}). An organized list of all the OPAs can be found on the  Gravitational-wave  Candidate  Event  Database (GraceDB, \citealt{followupprediction}). Although OPAs do not provide the raw GW data for a given event, they are useful in determining whether EM follow-up is warranted. To supplement the predictions available on GraceDB, \cite{GWSkynet} introduced a ``real-versus-noise" binary classifier, known as \texttt{GWSkyNet}, which leverages OPAs to inform potential EM follow-up seconds after the OPA is published. \texttt{GWSkyNet} achieves a test set accuracy of 93.5\% and correctly predicts 37 of the 40 O3a events published in the second Gravitational-Wave Transient Catalog (GWTC-2; \citealt{GWTC-2}), before the publication of this catalog. This binary classifier demonstrates the great potential for applications of machine learning to classification of GW candidates in real-time from the information provided in OPAs. These positive results encourage the implementation of a new extension to now have a multi-class classifier.

Here we present \texttt{GWSkyNet-Multi}, our new multi-class classifier which is an extension of the original \texttt{GWSkyNet}, and assess its performance. In contrast with the previous binary classification, our novel multi-class classifier categorizes GW candidates into their possible progenitors: (1) binary black hole (BBH) mergers, (2) mergers involving a neutron star (NS) (either NS+NS or NS+BH), or (3) non-astrophysical glitches. \texttt{GWSkyNet-Multi} empowers EM astronomers to make decisions on whether to follow-up an event informed by predictions of the nature of the source that supplement the information available in OPAs (\cite{MLinference}; LIGO/Virgo Public Alerts User Guide\footnote{\url{https://emfollow.docs.ligo.org/userguide/}}). \texttt{GWSkyNet-Multi} is a sequence of three convolutional neural network (CNN) one-versus-all classifiers. As a CNN, \texttt{GWSkyNet-Multi} distinguishes itself from other prediction methods such as \cite{EM_counterpart_prediction} and \citealt{Essick:2017jio} (\texttt{skymap\_statistics}\footnote{\href{https://github.com/reedessick/skymap\_statistics}{https://github.com/reedessick/skymap\_statistics}}) the latter of which employed mutual information distance of 2D GW skymaps to distinguish between signal and noise during O2. By running all of our models on a given event, we obtain three \textit{scores} quantifying the possibility that a source belongs to each class. Furthermore, using a hierarchical scheme, a refined prediction is obtained. We present this hierarchical scheme in  Figure \ref{fig: flowchart}. 

All \texttt{GWSkyNet-Multi} and \texttt{GWSkyNet} models, and scripts required to make predictions for a given event, are made publicly available.\footnote{\url{https://github.com/GWML/GWSkyNet}}.

The outline of this paper is as follows. In Section \ref{sec: methods}, we describe the new data set used for training and testing our new classifier and describe \texttt{GWSkyNet-Multi}. In Section \ref{sec: results}, we asses the performance of each individual model and the overarching hierarchical classifier. We further make predictions for sources from the LVC O3 observing run. Section \ref{sec: discussion} describes the performance and interpretability of our models. In Section \ref{sec: conclusion} we summarize our findings.

\begin{figure*}[!t]
    \captionsetup{width=\textwidth}
	\caption{\textbf{One-versus-all classifier architecture}. All three one-versus-all models used the architecture shown here. Each model takes as inputs a skymap image, three volume-projected images, the detector network status, mean and maximum distances, logarithms of two Bayes factors, and a normalization factor for each image. The shape of the inputs is given in parentheses, where N is the size of the training set. SeparableConv2D (SepConv2D) layers are described by the dimensions of the kernel and the number of filters. MaxPool layers are described only by the dimensions of the kernel. Finally, the numbers in the Dense layers indicate the width of that layer. 
	\label{fig: architecture}}{
   \includegraphics[width=\textwidth]{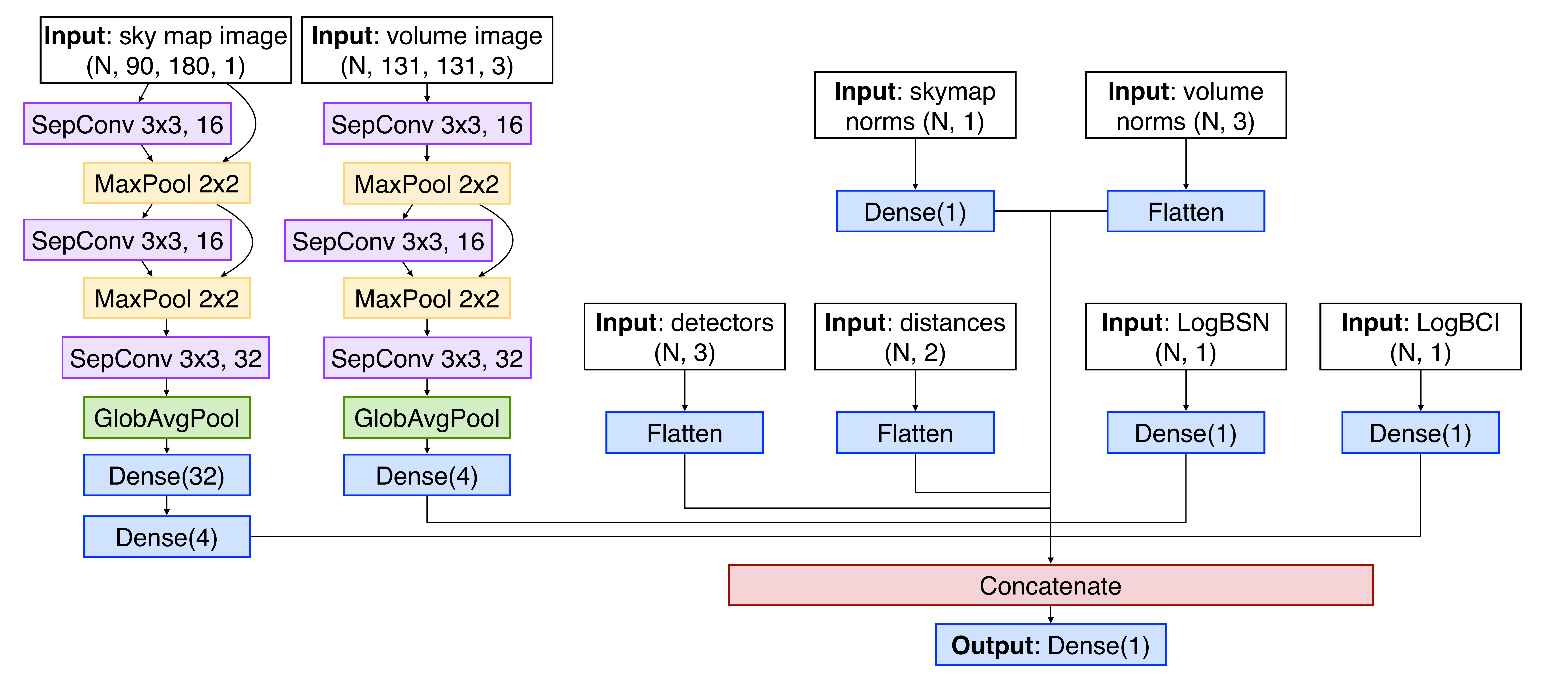}}
\end{figure*}

\section{Methods}\label{sec: methods}
We begin by describing the data used to train and test \texttt{GWSkyNet-Multi}. We then discuss the model's architecture and the creation of the multi-class classifier.

\subsection{Data Set}
The noise events in the data set consist of 1267 glitches from the first two observing runs of Advanced LIGO and Virgo, identified in~\cite{GWSkynet} using catalogs of noise transients~\citep{Blips,GravitySpy} and GW candidates from the second Open Gravitational-wave Catalog~\citep{2-OGC}. To construct a sufficiently large and balanced data set, we use gravitational waveform models from the \texttt{LALSuite} package (\citealt{lalsuite}) to simulate 1000 GW events for each astrophysical source type: BBH, BNS mergers, and NSBH mergers. The noise realization for the simulated events is achieved by injecting the GW waveforms into Gaussian noise coloured with publicly available power sprectral densities (PSDs) from O1, O2, and O3 (\citealt{noise_for_sumulated_events}). BH masses are drawn from astrophysically-motivated mass distributions described in \cite{GWTC-2-populations}. We choose the ``power law + peak'' mass model and its corresponding mass-ratio distribution, using the median posterior values for the parameters of the model. NS masses follow a uniform distribution in the range $[1, 3]~M_\odot$. Spins are restricted to be along the direction of orbital angular momentum, with spin magnitudes constrained to be $\leq 0.99$ for BH and $\leq 0.05$ for NS.

The full data set contains 4267 events. We use 81\% of this set for training, 9\% for validation, and 10\% is reserved for the test set.

\subsection{Developing a Multi-Class Classifier}\label{sec: developing a multi-classifier}

In order to make a prediction from an event, \texttt{GWSkyNet-Multi} uses the following inputs derived from the OPA FITS file generated by \texttt{BAYESTAR} (\citealt{BAYESTAR}):
\begin{enumerate}
    \item A skymap image;
    \item Three volume-projected images;
    \item A list of detectors observing at the time of detection;
    \item Estimated mean and maximum distance to the source of the GW;
    \item Four normalization factors (one for the skymap image, and one for each of the three volume-projected images);
    \item The logarithm of the Bayes factors for the signal versus noise hypothesis (LogBSN) and the coherence versus incoherence hypothesis (LogBCI).
\end{enumerate}

\begin{figure}[H]
	\centering
	\includegraphics[width=0.5\textwidth]{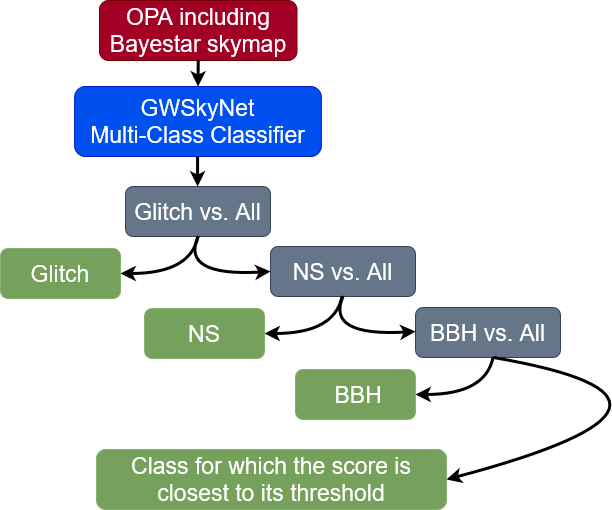}
	\caption{
	\textbf{Flow  chart  of  the  hierarchical  multi-class classifier.} Given an OPA candidate, we run all three classifiers to obtain scores for each model. All scores are returned to the user. To obtain a single prediction, we use a hierarchical scheme: If the glitch-versus-all score is above its threshold, we classify the event as a glitch. If not, we check if the NS-versus-all score is above its threshold, and if so, we classify the event as containing a NS. If the event is neither a glitch nor a NS according to these criteria, we check if the BBH-versus-all score is above its threshold, and classify the event as BBH if so. If no model's score is above its corresponding threshold, we select the model for which the score was closest to the threshold. We discuss the choice of threshold in Section 3, but note that the user may set their own thresholds.}
	\label{fig: flowchart}
\end{figure}

To generate a multi-class classifier, we use a sequence of three one-versus-all classifiers. All one-versus-all models have the same architecture, shown in Figure \ref{fig: architecture}. Each model is used to classify an event as belonging to a given class versus all remaining classes, \textit{e.g.}, BBH-versus-all classifies events as either a BBH merger or anything which is not a BBH merger. We thus obtain three models: BBH-versus-all, NS-versus-all, and glitch-versus-all. We tune these models by varying the ``learning rate", a scalar used to vary the magnitude of the step size in a gradient descent algorithm, as well as the ``batch size", the number of examples used in each step during training. To find the hyperparameters leading to the best models for each class, we use \texttt{Hparams}\footnote{\href{https://pypi.org/project/hparams/}{https://pypi.org/project/hparams/}} to explore hyperparameter-space with learning rates ranging from $10^{-5}$ to $10^{-1}$, and batch sizes ranging from 5 to 110. Once trained, all three models are applied to an event, and so we obtain a score quantifying the possibility that a candidate belongs to each class. Finally, to return a single prediction for our analysis, we use the hierarchical scheme shown in Figure \ref{fig: flowchart}.

For a given input, if a one-versus-all classifier returns a score above the user-set prediction threshold, the candidate is classified as a member of that particular class. If the score is below the threshold, we proceed to the next classifier in the hierarchical scheme. If none of the scores are above the models' thresholds, we classify the event based on which score was closest to its threshold.

We run models in order of best-performing to worst, based on both test set accuracies and $F_1$ scores: 

\begin{equation}\label{eq: accuracy}
\text{accuracy} = \frac{\mathrm{TP} + \mathrm{TN}}{\mathrm{TP} + \mathrm{TN} + \mathrm{FP} + \mathrm{FN}},
\end{equation}

\begin{equation}\label{eq: f1 score}
F_1~\mathrm{score} = \frac{\mathrm{TP}}{\mathrm{TP} + \frac{1}{2}(\mathrm{FP} + \mathrm{FN})},
\end{equation}

\noindent where TP is the number of true positive predictions, TN is the number of true negatives, FP is the number of false positives, and FN is the number of false negatives. We primarily focus on the accuracy of each model, however we also restrict ourselves to using models with high F$_1$ scores ($\geq 0.95$) as they predict a sufficient amount of TPs.

We employ a hierarchical scheme to ensure that the prediction of our best-performing one-versus-all model is given priority over the predictions of the less trustworthy models. As seen in the next section, our best-performing classifier is glitch-versus-all, followed by NS-versus-all, and finally BBH-versus-all. However, note that~\texttt{GWSkyNet-Multi} provides the user with both the hierarchical prediction and the  score for each individual classifier, to maximally inform EM follow-up.

\section{Performance \& Predictions}\label{sec: results}
Next, we  present the performance of each one-versus-all classifier, as well as the predictions of the multi-class classifier, for candidates in LIGO--Virgo's O3 run.

\begin{figure}[ht]
	\centering
	\includegraphics[width=0.47\textwidth]{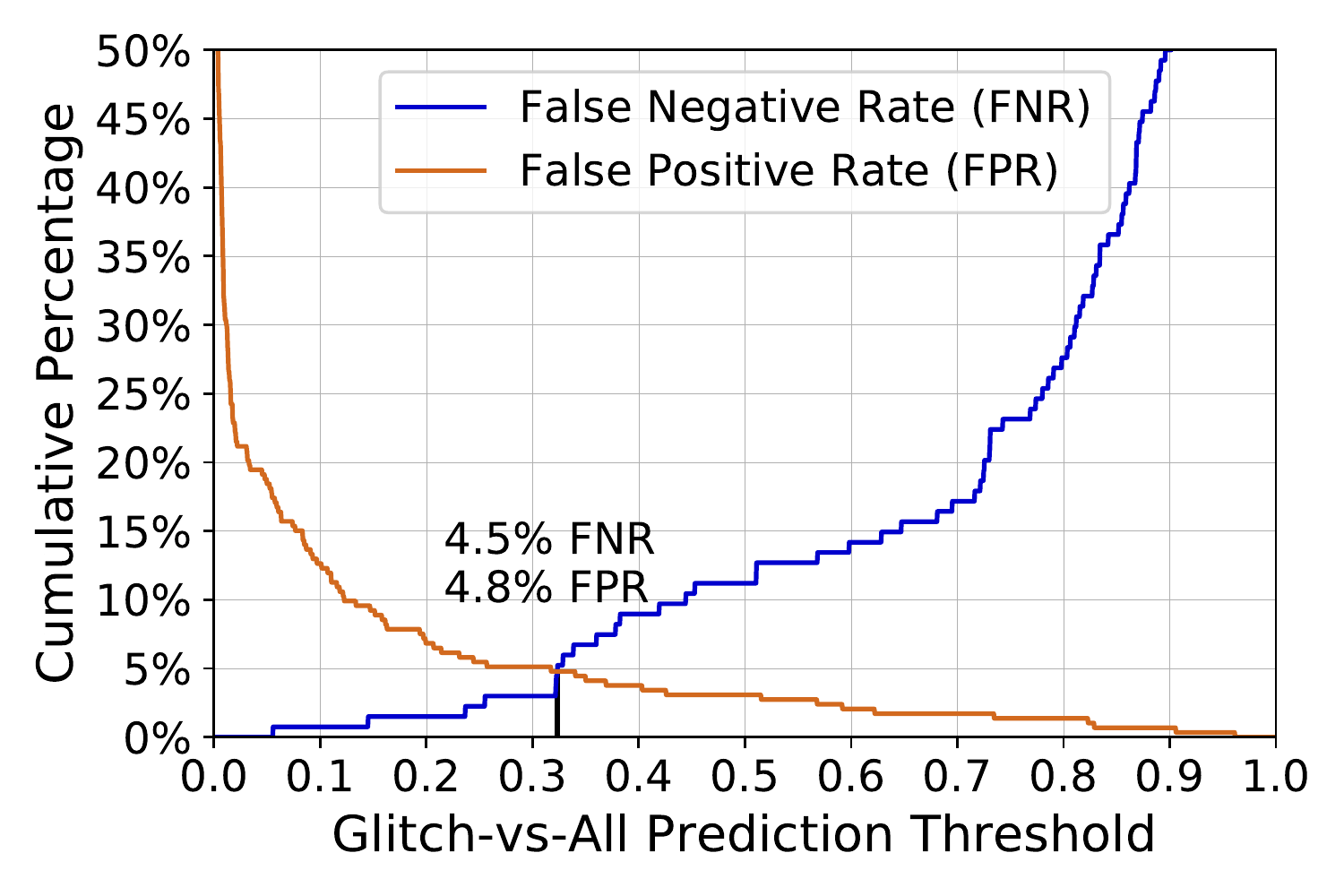}
	\includegraphics[width=0.47\textwidth]{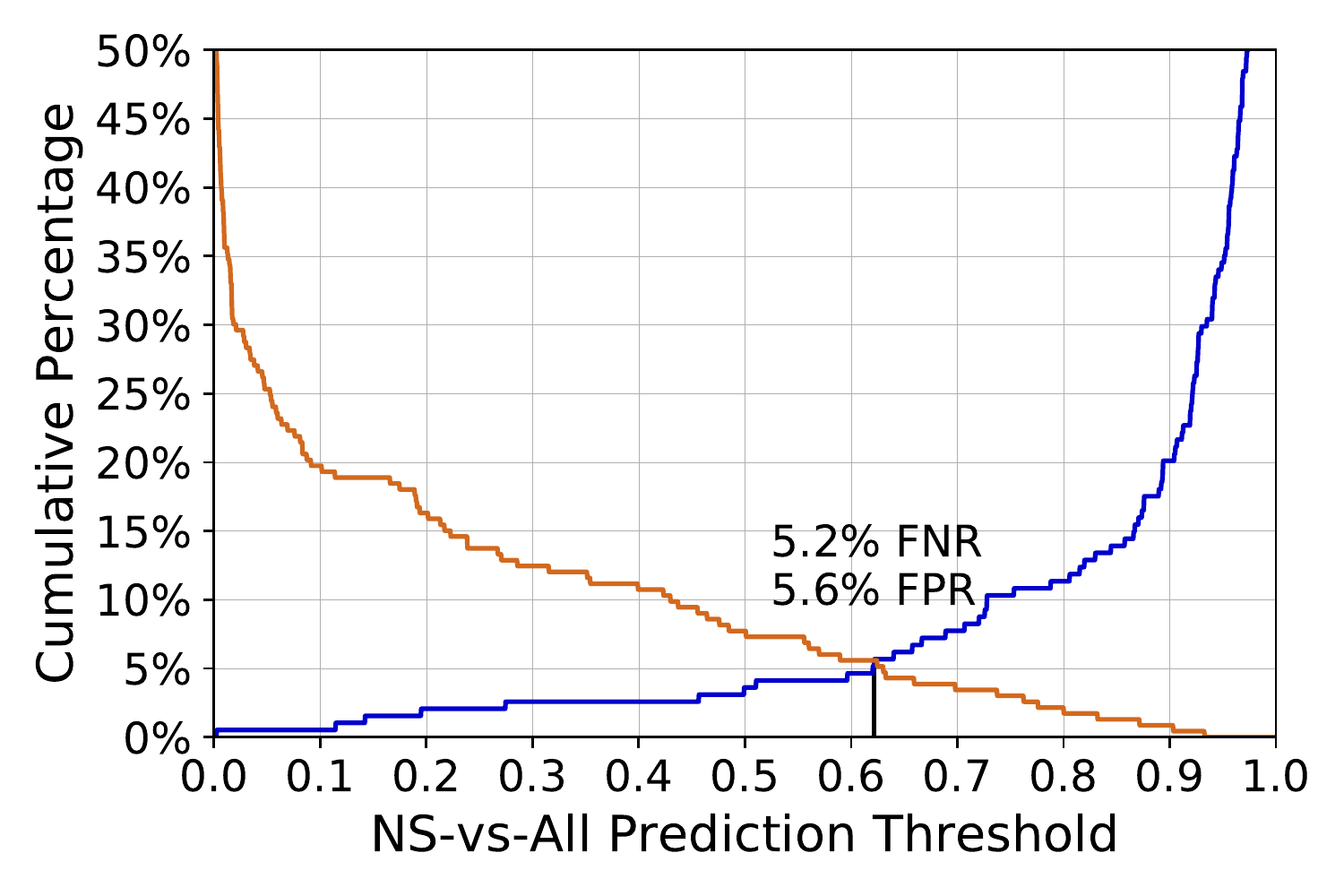}
	\includegraphics[width=0.47\textwidth]{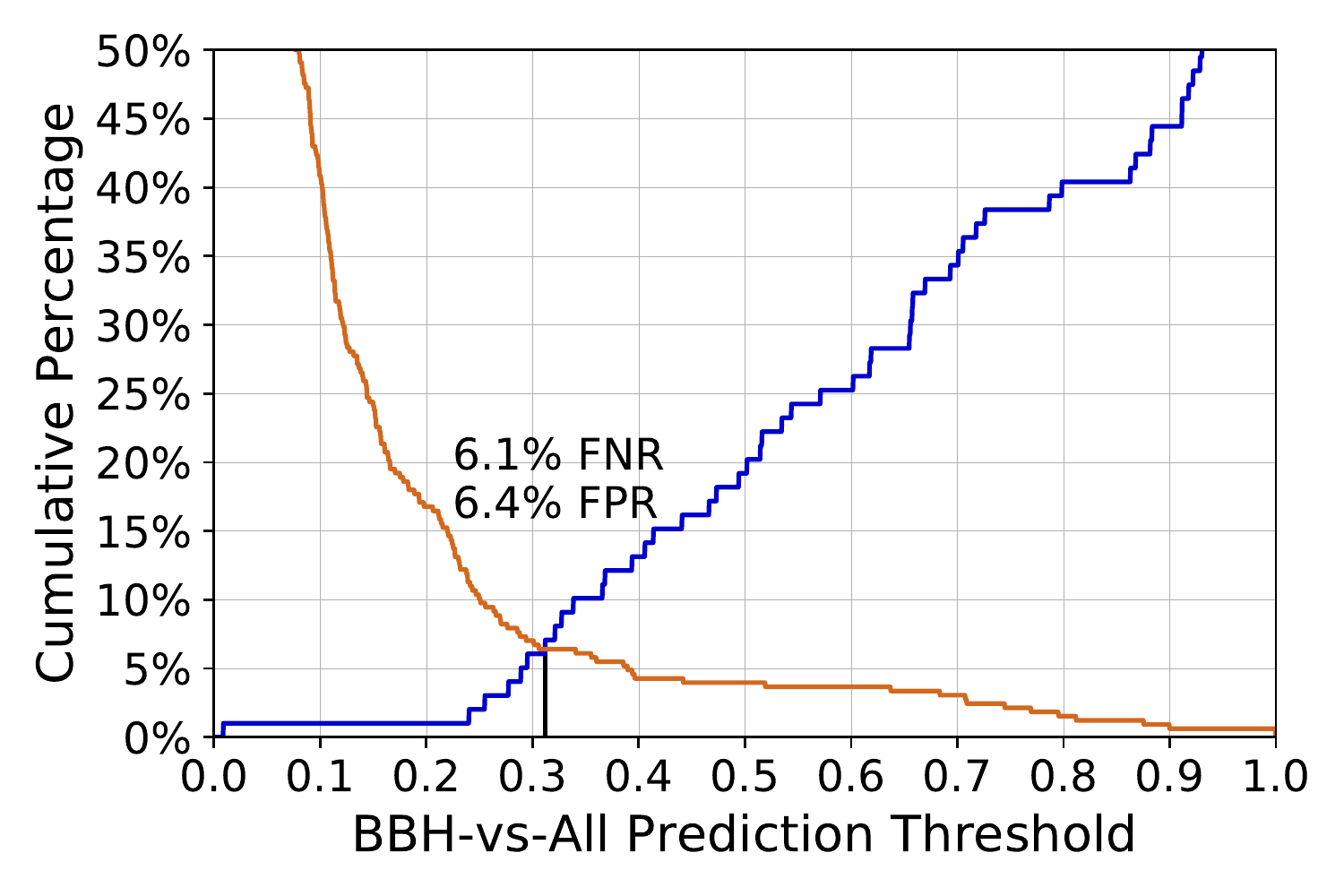}
	\caption{\textbf{Prediction threshold plots.} Along the x-axis is the one-versus-all classifier's threshold and along the y-axis is the cumulative percentage of events which are classified. In order from top panel to bottom panel, we show the glitch-versus-all, NS-versus-all, and BBH-versus-all threshold plots. The threshold used for the predictions presented in this paper is where the false negative rate (FNR), in blue, and the false positive rate (FPR), in orange, intersect, marked by a black vertical line. For the glitch-versus-all classifier, the intersection is at 32.3\%  score, for the NS-versus-all classifier 62.2\%  score and for the BBH-versus-all 31.2\%.}
	\label{fig: glitch threshold}
\end{figure}

Figure \ref{fig: glitch threshold} shows the FNR and FPR of the predictions on the test set made by each of the one-versus-all classifiers as a function of the threshold. We find that the best performance for each one-versus-all model occurs when the prediction threshold is set to the intersection between the false negative rate (FNR) and the false positive rate (FPR). However, the user may change this threshold in \texttt{GWSkyNet-Multi} to suit their preferences for the FPR and FNR of the classifiers. The accuracy and $F_1$ score for each one-versus-all classifier, applied to the 427 examples in the test set using the intersection threshold, are presented in Table \ref{tab: test set results}.

\begin{deluxetable}{lccc}[!ht]
\tablenum{1}
\caption{\textbf{Results of the one-versus-all classifiers on the test set.} Here, we present the threshold, accuracy, and $F_1$ score for each model. The corresponding confusion matrices are presented in Figure \ref{fig: confusion matrices} (Appendix~\ref{app:confusionmatrices}).\label{tab: test set results}}
\tablewidth{0pt}
\tablehead{\makecell{\textbf{Classifier}} & \makecell{\textbf{Threshold}} & \makecell{\textbf{Accuracy}} & \makecell{\textbf{$F_1$ score}} }
\startdata
Glitch-versus-all & 32.3\% & 95.1\% & 0.96 \\
NS-versus-all & 62.2\% & 94.4\% & 0.95 \\
BBH-versus-all & 31.2\% & 93.7\% & 0.96
\enddata
\end{deluxetable}
\vspace{-4 em}

The glitch-versus-all, NS-versus-all, and BBH-versus-all classifiers achieve accuracies of 95.1\%, 94.4\%, and 93.7\%, and $F_1$ scores of 0.96, 0.95, and 0.96, respectively. To calculate the accuracy for the overall multi-class classifier, the three one-versus-all models are run sequentially on the test set as described in Section \ref{sec: developing a multi-classifier}. We obtain an overall test set accuracy of 93.4\% with this hierarchical scheme. We also obtain multi-class predictions for O3a events with this hierarchical method. The hierarchical classifications and individual one-versus-all  scores are reported in Table \ref{tab: O3a results}. Finally, although the refined classifications for O3b events have not yet been published, we present the predictions of \texttt{GWSkyNet-Multi} in Table \ref{tab: O3b results}.

\begin{deluxetable*}{lcccccccc}[!ht]
\tablenum{2}
\caption{\textbf{O3a predictions}. Columns are: the OPA candidate names for O3a events with \texttt{BAYESTAR}-generated skymaps, and their labels according to GWTC-2, GraceDB, \texttt{GWSkyNet}, and \texttt{GWSkyNet-Multi} (individual and hierarchical scores). The GWTC-2 labels are obtained assuming a maximum neutron star mass of 3~$M_\odot$. The GraceDB labels are obtained by selecting the class with the highest probability. We highlight hierarchical predictions which do not match those of GWTC-2 in bold. Finally, on one side, ambiguous events whose prediction scores were all below their respective thresholds are marked with an asterisk (*). In addition, the ambiguous cases where more than one prediction score is above the classifier's threshold, are marked with a dagger (\vdag) \label{tab: O3a results}}
\tablewidth{0pt}
\tablehead{\makecell{\textbf{Candidate} \\ \textbf{Name}} & \makecell{\textbf{GWTC-2} \\ \textbf{Label}} & \makecell{\textbf{GraceDB} \\ \textbf{Label}} & \makecell{\texttt{GWSkyNet} \\ \texttt{Binary}} & \makecell{\textbf{Glitch} \\ \textbf{Score (\%)}} & \makecell{\textbf{NS} \\ \textbf{Score (\%)}} & \makecell{\textbf{BBH} \\ \textbf{Score (\%)}} &  \makecell{\textbf{Hierarchical} \\ \textbf{Prediction}}}
\startdata
S190405ar & Glitch & RETRACTED & Glitch & 30.3 & 4.3 & 1.2 & Glitch* \\
S190408an & BBH & BBH & Real & 0.1 & 0.1 & 99.9 & BBH \\
S190412m & BBH & BBH & Real & 0.1 & 4.2 & 94.1 & BBH \\
S190421ar & BBH & BBH & Real & 23.8 & 0 & 100 & BBH \\
S190425z & BNS & BNS & Real & 89.5 & 99.9 & 0 & \textbf{Glitch}\vdag \\
S190426c & NSBH & BNS & Real & 0.1 & 97.2 & 21.2 & NS \\
S190503bf & BBH & BBH & Real & 0.2 & 30.6 & 40.4 & BBH \\
S190510g & Glitch & Terrestrial & Glitch & 38.7 & 1 & 3.8 & Glitch \\
S190512at & BBH & BBH & Real & 0.1 & 0 & 99.7 & BBH \\
S190513bm & BBH & BBH & Real & 0.1 & 0 & 100 & BBH \\
S190517h & BBH & BBH & Real & 0.1 & 0 & 100 & BBH \\
S190518bb & Glitch & RETRACTED & Glitch & 99.9 & 96.8 & 4.2 & Glitch\vdag \\
S190519bj & BBH & BBH & Real & 0.1 & 0 & 100 & BBH \\
S190521g & BBH & BBH & Real & 0.1 & 1.3 & 71.2 & BBH \\
S190521r & BBH & BBH & Real & 0.4 & 1.6 & 99.3 & BBH \\
S190524q & Glitch & RETRACTED & Glitch & 78.9 & 74.7 & 0.9 & Glitch\vdag \\
S190602aq & BBH & BBH & Real & 0.1 & 6.4 & 91.9 & BBH \\
S190630ag & BBH & BBH & Real & 72.3 & 0 & 76.2 & \textbf{Glitch}\vdag \\
S190701ah & BBH & BBH & Real & 0.1 & 0.2 & 97.7 & BBH \\
S190706ai & BBH & BBH & Real & 0.1 & 0 & 100 & BBH \\
S190707q & BBH & BBH & Real & 1 & 18.2 & 83.5 & BBH \\
S190718y & Glitch & Terrestrial & Glitch & 88.7 & 0.3 & 19.4 & Glitch \\
S190720a & BBH & BBH & Real & 3.2 & 3.3 & 96.7 & BBH \\
S190727h & BBH & BBH & Real & 0.2 & 0.1 & 98.4 & BBH \\
S190728q & BBH & BBH & Real & 0.1 & 15.6 & 90.8 & BBH \\
S190808ae & Glitch & RETRACTED & Glitch & 47.7 & 98.7 & 0.1 & Glitch \\
S190814bv & NSBH & NSBH & Real & 0.1 & 94.1 & 45.2 & NS\vdag\\
S190816i & Glitch & RETRACTED & Real & 1.4 & 87 & 18.2 & \textbf{NS} \\
S190822c & Glitch & RETRACTED & Glitch & 60.8 & 100 & 9.2 & Glitch\vdag \\
S190828j & BBH & BBH & Real & 0.1 & 0 & 100 & BBH \\
S190828l & BBH & BBH & Real & 0.1 & 0 & 99.9 & BBH \\
S190829u & Glitch & RETRACTED & Real & 70.6 & 63.4 & 0 & Glitch\vdag \\
S190901ap & Glitch & BNS & Glitch & 76.6 & 90.1 & 0 & Glitch\vdag \\
S190910d & Glitch & NSBH & Glitch & 45.6 & 0.5 & 0 & Glitch \\
S190910h & Glitch & BNS & Glitch & 99 & 0 & 0 & Glitch \\
S190915ak & BBH & BBH & Real & 0.1 & 0 & 99.9 & BBH \\
S190923y & Glitch & NSBH & Real & 16.9 & 63.6 & 3.9 &  \textbf{NS}\\
S190924h & BBH & MassGap & Real & 0.1 & 30.9 & 58.6 & BBH \\
S190930s & BBH & MassGap & Real & 29.2 & 32.2 & 60.4 & BBH \\
S190930t & Glitch & NSBH & Glitch & 99.4 & 0 & 0 & Glitch
\enddata
\tablecomments{GW190413, GW190413, GW190424, GW190514, GW190527, GW190620, GW190708, GW190719, GW190731, GW190803, GW190803, GW190909, GW190910, GW190929 were identified in the GWTC-2 catalog however, they did not produce OPAs at the time of detection and thus \texttt{GWSkyNet} does not make predictions on these events.}
\end{deluxetable*}

\begin{deluxetable*}{lccccccc}[!ht]
\tablenum{3}
\caption{\textbf{O3b predictions}. Columns are: the OPA event names for O3b events with \texttt{BAYESTAR}-generated skymaps, and their labels according to GraceDB, \texttt{GWSkyNet}, and \texttt{GWSkyNet-Multi} (individual and hierarchical scores). Ambiguous events whose prediction scores were all below their respective thresholds are marked with an asterisk(*). In addition, the ambiguous cases where more than one prediction score is above the classifier's threshold, are marked with a dagger (\vdag).\label{tab: O3b results}}
\tablewidth{0pt}
\tablehead{\makecell{\textbf{Candidate} \\ \textbf{Name}} & \makecell{\textbf{GraceDB} \\ \textbf{Label}} & \makecell{\texttt{GWSkyNet} \\ \texttt{Binary}} & \makecell{\textbf{Glitch} \\ \textbf{Score (\%)}} & \makecell{\textbf{NS} \\ \textbf{Score (\%)}} & \makecell{\textbf{BBH} \\ \textbf{Score (\%)}} & \makecell{\textbf{Hierarchical} \\ \textbf{Prediction}}}
\startdata
S191105e & BBH & Glitch & 0.1 & 0 & 98.7 & BBH \\
S191109d & BBH & Real & 15.6 & 0 & 100 & BBH \\
S191110x & RETRACTED & Real & 90.2 & 86.6 & 17.9 & Glitch\vdag \\
S191117j & RETRACTED & Real & 100 & 70.2 & 6.5 & Glitch\vdag \\
S191120aj & RETRACTED & Glitch & 92.8 & 34 & 0.1 & Glitch \\
S191120at & RETRACTED & Real & 99.4 & 98.7 & 27 & Glitch\vdag \\
S191124be & RETRACTED & Glitch & 95.1 & 91.2 & 5.6 & Glitch\vdag \\
S191129u & BBH & Real & 0.8 & 0.5 & 77.1 & BBH \\
S191204r & BBH & Real & 0.1 & 3.1 & 84.1 & BBH \\
S191205ah & NSBH & Glitch & 50.4 & 97 & 0.1 & Glitch\vdag \\
S191212q & RETRACTED & Real & 49.8 & 100 & 1.9 & Glitch\vdag \\
S191213ai & RETRACTED & Glitch & 98.3 & 92.1 & 0 & Glitch\vdag \\
S191213g & BNS & Real & 3 & 2.4 & 2.7 & BBH* \\
S191215w & BBH & Real & 0.1 & 0 & 100 & BBH \\
S191216ap & BBH & Real & 1.7 & 83.4 & 43.2 & NS\vdag \\
S191220af & RETRACTED & Real & 5.6 & 99.8 & 4.7 & NS \\
S191222n & BBH & Real & 16.5 & 0 & 18.8 & BBH* \\
S191225aq & RETRACTED & Real & 0.2 & 98.5 & 8.7 & NS \\
S200105ae & Terrestrial & Real & 54.9 & 89.5 & 0 & Glitch\vdag \\
S200106au & RETRACTED & Real & 0.7 & 84.8 & 9.9 & NS \\
S200106av & RETRACTED & Real & 32.7 & 58.9 & 2.6 & Glitch \\
S200108v & RETRACTED & Glitch & 98.4 & 25.7 & 10.1 & Glitch \\
S200112r & BBH & Real & 76.2 & 0 & 44.4 & Glitch\vdag \\
S200115j & MassGap & Real & 0.2 & 27.5 & 27.8 & BBH* \\
S200116ah & RETRACTED & Real & 99 & 93.3 & 3.8 & Glitch\vdag\\
S200128d & BBH & Real & 17.7 & 0 & 100 & BBH \\
S200129m & BBH & Real & 0.1 & 0.9 & 98.2 & BBH \\
S200208q & BBH & Real & 0.1 & 0 & 100 & BBH \\
S200213t & BNS & Glitch & 53 & 99.8 & 2.8 & Glitch\vdag\\
S200219ac & BBH & Real & 0.1 & 0 & 99.9 & BBH \\
S200224ca & BBH & Real & 0 & 0 & 100 & BBH \\
S200225q & BBH & Real & 0.8 & 0 & 98.9 & BBH \\
S200302c & BBH & Glitch & 95.2 & 0 & 0.7 & Glitch \\
S200303ba & RETRACTED & Glitch & 63.4 & 0 & 85.1 & Glitch\vdag\\
S200308e & RETRACTED & Real & 98.8 & 75.6 & 4.7 & Glitch\vdag \\
S200311bg & BBH & Real & 0 & 1.7 & 95.8 & BBH \\
S200316bj & MassGap & Real & 0.2 & 0.3 & 99.3 & BBH
\enddata
\tablecomments{S200105ae and S200115j were determined to be a real events in \cite{Real_events}.}
\end{deluxetable*}
\newpage
\clearpage
\section{Discussion}\label{sec: discussion}
To evaluate our model, we compare our O3 predictions with GWTC-2 (\citealt{GWTC-2}), GraceDB, and \texttt{GWSkyNet}. We then characterize the interpretability of our model using Gradient-weighted Class Activation Mapping (Grad-CAM).

\subsection{O3 Predictions Discussion}\label{sec: O3 predictions discussion}
During the O3a observing run, 40 gravitational wave candidates were made available as OPAs containing \texttt{BAYESTAR}-generated skymaps. By comparing the hierarchical predictions with those of GWTC-2, we find that 
\texttt{GWSkyNet-Multi} correctly classifies 36 of the 40 events, including 6/7 of the retracted events. The four misclassified events are further discussed in the following subsections. Furthermore, in comparison to the \texttt{GWSkyNet} classifier, the new model shows more promising results on retracted events. When comparing the 8 retracted events for which \texttt{GWSkyNet} and \texttt{GWSkyNet-Multi} have contradicting predictions. \texttt{GWSkyNet} correctly classifies 1 as a glitch whereas \texttt{GWSkyNet-Multi} correctly classifies 7 as glitches. On non-retracted events, we achieve comparable accuracy to \texttt{GWSkyNet} and GraceDB which correctly predicted 30/33 and 28/33, respectively (\texttt{GWSkyNet-Multi} obtains 29/33).

\subsubsection{S190425z}\label{S190425z}

We misclassify S190425z (now GW190425) as a glitch, while the refined analysis presented in \cite{GW190425} determines that it was a binary neutron star merger. Although S190425z was a source of interest and many electromagnetic observers attempted to find a counterpart (\citealt{GW190425}), it was also one of the most poorly-localized events in O3. The 90\% confidence region of this event was over 7000~$\mathrm{deg}^2$ as shown in Figure \ref{fig: skymap}. Classification of this event will be challenging with any model. In such cases where localization is poor, it can be useful to examine the scores of each individual classifier rather than relying solely on the hierarchical classifier. In the case of S190425z, we see that the glitch-versus-all classifier assigns a score of 89.5\% and the NS-versus-all classifier 99.9\%. The misclassification is thus entirely due to the fact that the glitch-versus-all is placed before the NS-versus-all in the hierarchical scheme.

\subsubsection{S190630ag}\label{S190630ag}

Similarly to the previous event, S190630ag is misclassified as a glitch, when the event was instead a BBH merger. However, our BBH-versus-all classifier has a score of 76.2\%, which is above its threshold. The misclassification is thus again due to the hierarchical scheme, and shows why it is important to take all the one-versus-all scores into consideration.

\subsubsection{S190816i}
We misclassify S190816i as containing a NS, when it was instead a glitch. We note however that this source was initially classified as either a NSBH or Terrestrial on GraceDB with 83\% and 17\% probability, respectively, before being retracted. We further note the high false alarm rate (FAR) presented on GraceDB: 1 per 2.2067 years. The weak signal of this source, as evident in the FAR, is likely responsible for the initial misclassification on GraceDB and the misclassification by our models.

\subsubsection{S190923y}
As with S190816i, we misclassify S190923y as containing a NS, when it was instead a glitch. Similar to this previous source, this source was also initially classified as either a NSBH or Terrestrial on GraceDB with 68\% and 32\% probability, respectively. We further note the high false alarm rate (FAR) presented on GraceDB: 1.5094 per year. 
\newline\newline

\subsection{Ambiguous Events}
There are four O3 events whose scores are all below their respective classifier's threshold, namely, S190405ar, S191213g, S191222n, S200115j. Of these events, the classifier score of S190405ar, S191222n, S200115j is highest in the correct class. We note that S200115j, which was labelled as MassGap by GraceDB, is classified with a near equal score by the NS-versus-all and BBH-versus-all classifiers. We further note that S191213g interestingly has low scores (less than 5\%) in each of the three classes, which suggests that this candidate may be a unique source which was not well characterized by the training set.

In addition, there are 21 events in O3 where more than one prediction score is above the classifier's threshold. For these events, including S190425z and S190630ag mentioned above in Sections \ref{S190425z} and \ref{S190630ag}, respectively, it is important to look at all the predictions scores in addition to the hierarchical prediction. 



\subsection{Interpretability}\label{sec: interpretability}

\begin{figure}[!ht]
	\centering
	\includegraphics[width=0.45\textwidth]{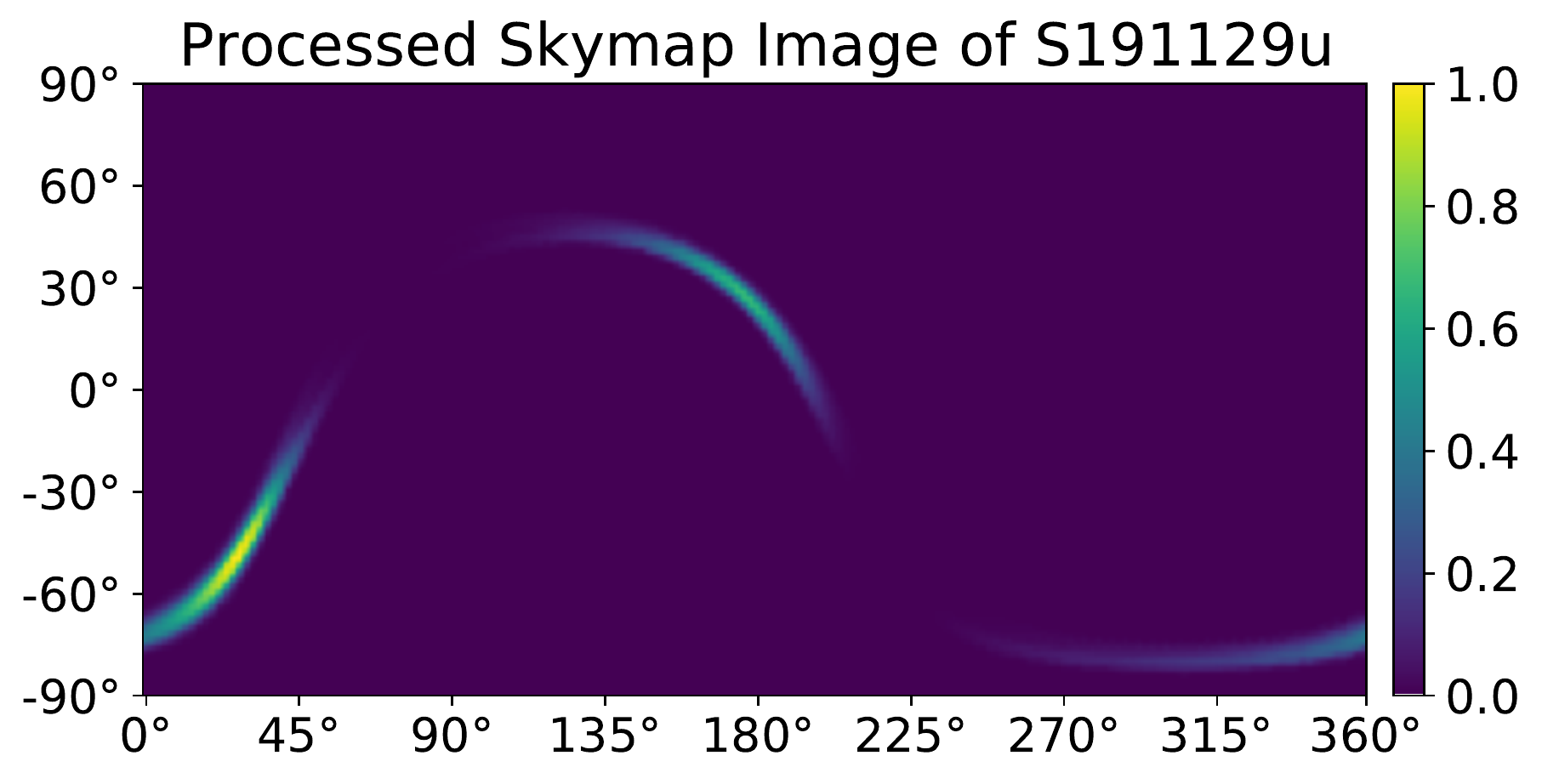}
	\includegraphics[width=0.45\textwidth]{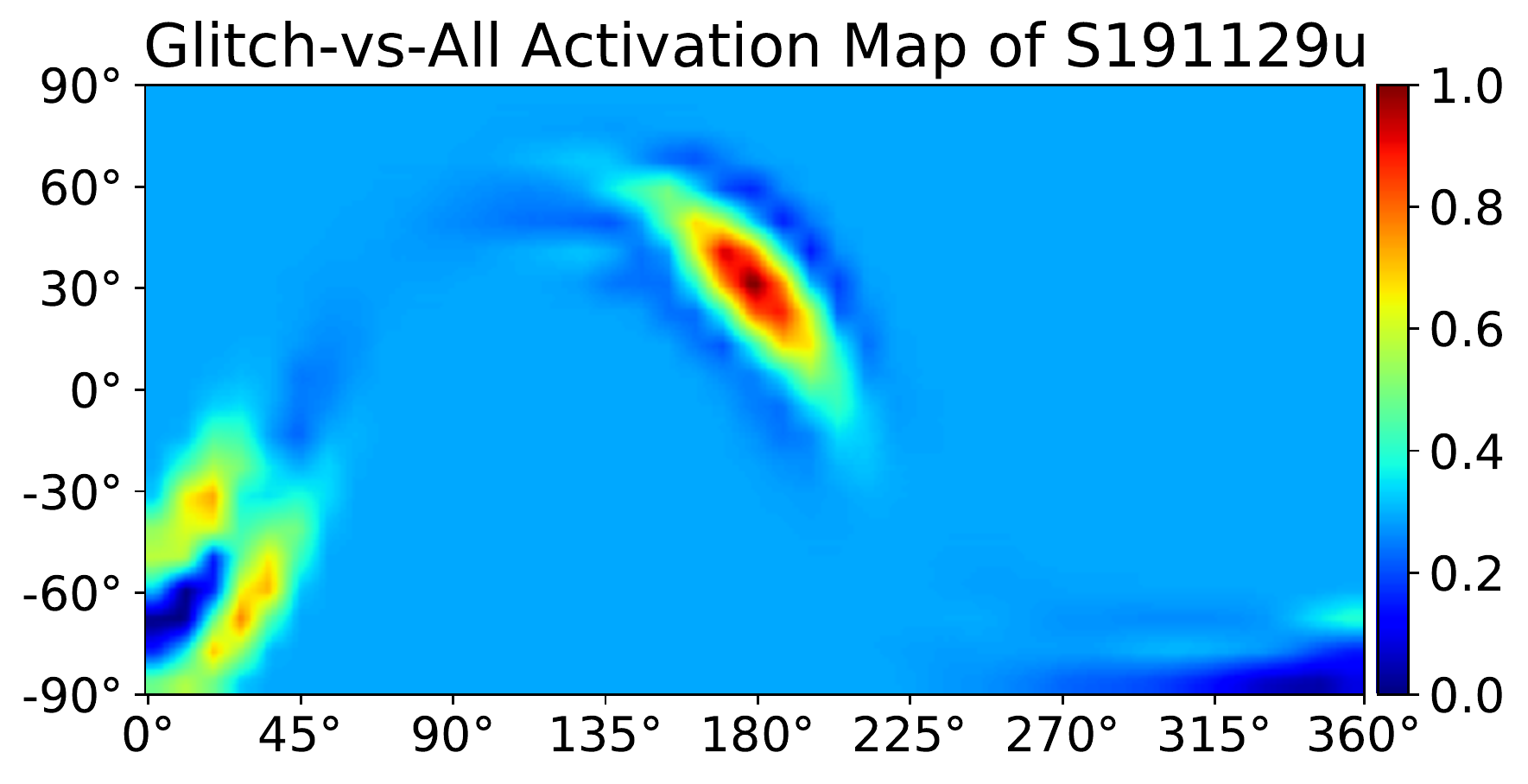}
	\includegraphics[width=0.45\textwidth]{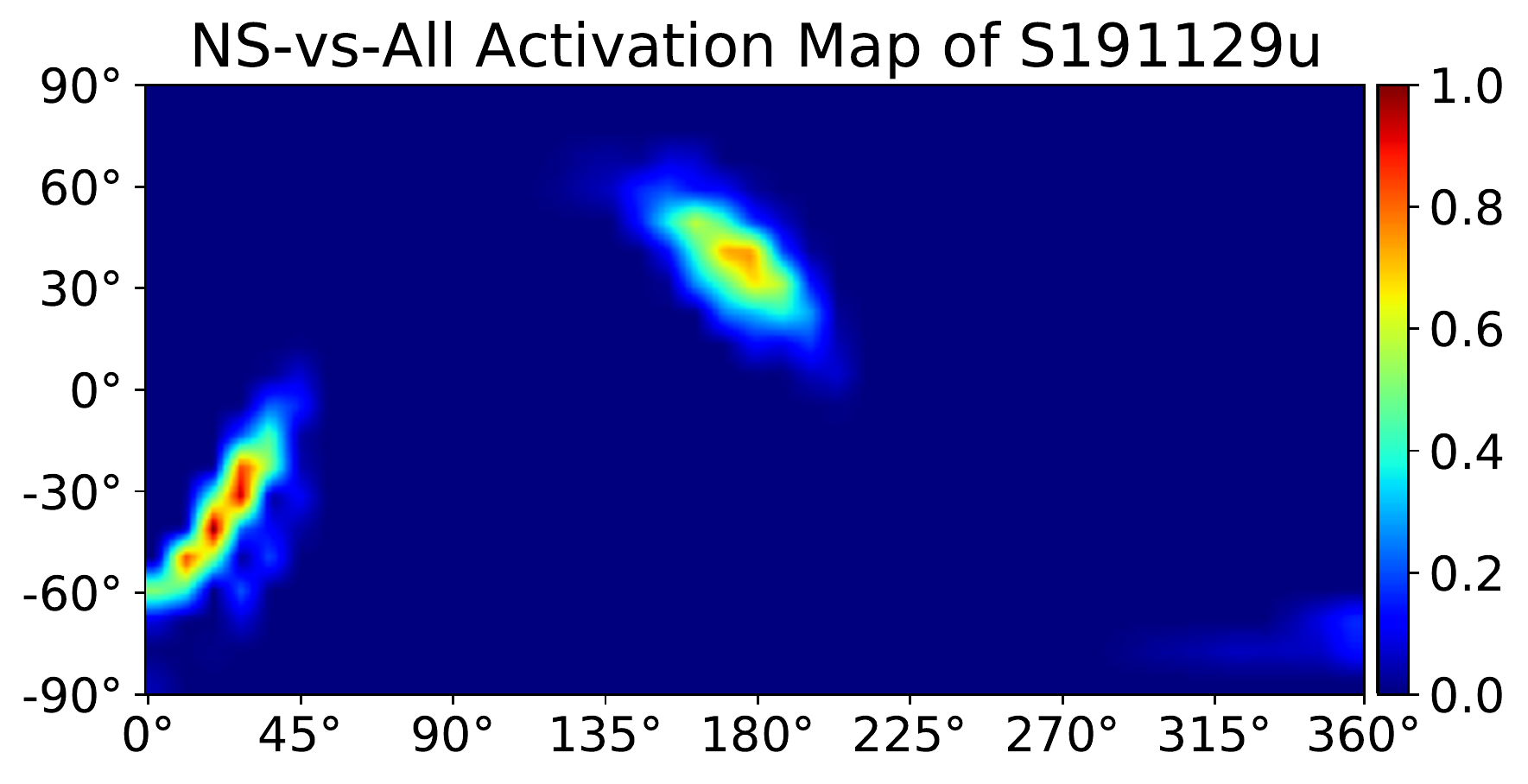}
	\includegraphics[width=0.45\textwidth]{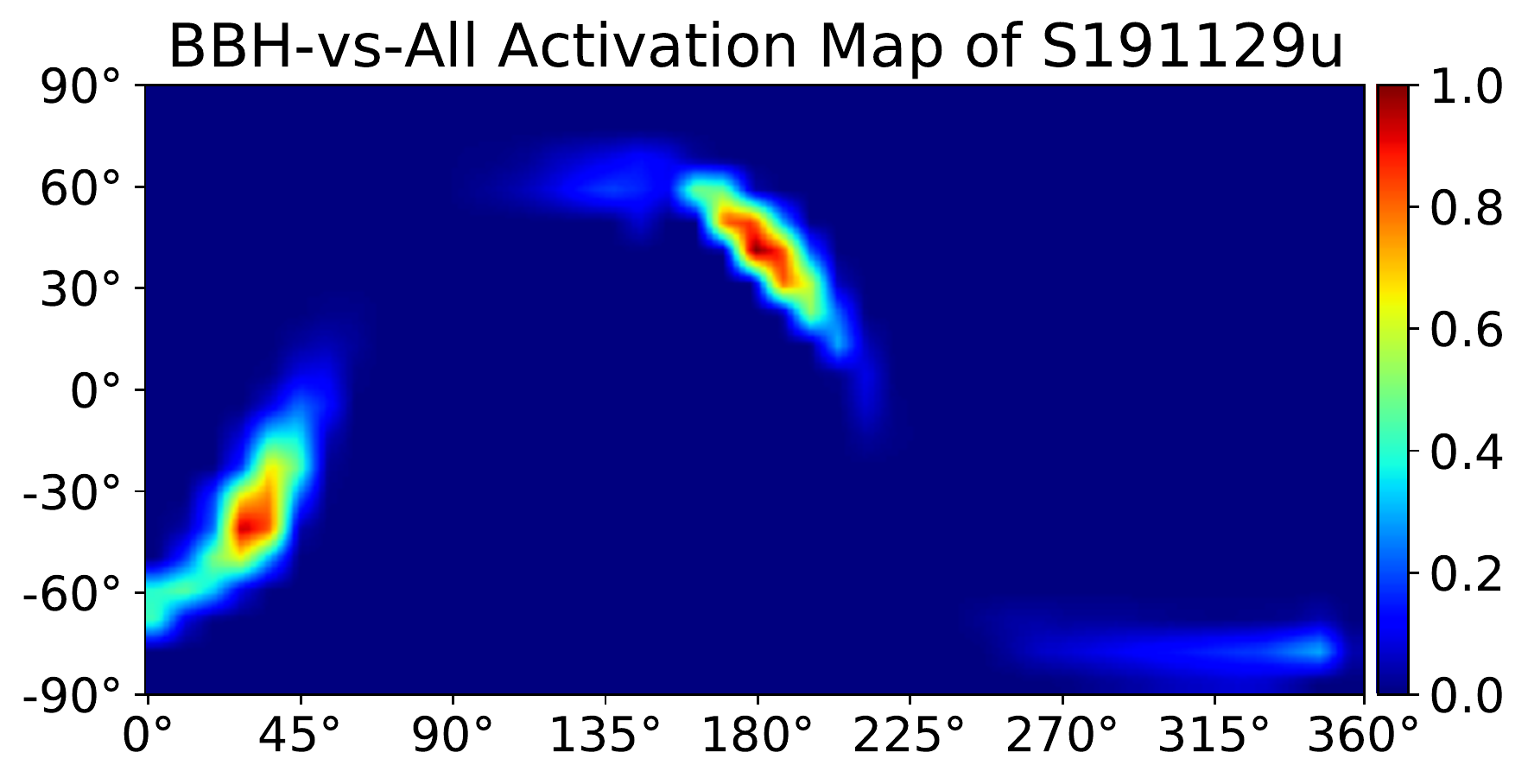}
	\caption{\textbf{Grad-CAM activation maps of S191129u.} \textit{Top panel:} Processed skymap of the S191129u localization region. \textit{Remaining panels:} Grad-CAM activation maps highlighting the portion of the skymap that the given models analyze most in order to make a prediction. A Grad-CAM activation map is shown for each of the three one-versus-all classifiers. This event is correctly classified by \texttt{GWSkyNet-Multi} as a BBH, with scores of 0.1\%, 15.6\%, and 90.8\% for glitch-versus-all, NS-versus-all, and BBH-versus-all, respectively. Activation maps are smoothed using a bilinear interpolator.}
	\label{fig: GradCAM S191129u}
\end{figure}

Intuitively, we expect that a given classifier would perform best when it focuses it's attention on the high-probability regions of the skymap. We tested this intuition using Gradient-weighted Class Activation-Mapping (Grad-CAM)\footnote{\href{https://github.com/ramprs/grad-cam/}{https://github.com/ramprs/grad-cam/}}. Grad-CAM makes use of the fact that convolutional layers preserve spatial information to create an ``activation map". Each map is produced using gradients of the convolutional layers calculated with respect to the feature map activations. From the activation map, regions of a given skymap which receive the most attention in making a given classification are identified (\citealt{Grad-CAM}). With Grad-CAM, we see that this intuition is generally correct. However, we caution that we observe a general trend in the glitch-versus-all classifier, where there is more attention on the low-probability regions than is present for the other classifiers (see Figure \ref{fig: GradCAM S191129u}). Despite this fact, the glitch-versus-all classifier performs best of all the classifiers.

\begin{figure*}[!ht]
    \subfloat[Processed skymap of the S19012at localization region.\label{fig: gradcam  a}]
    {\includegraphics[width=0.5\textwidth]{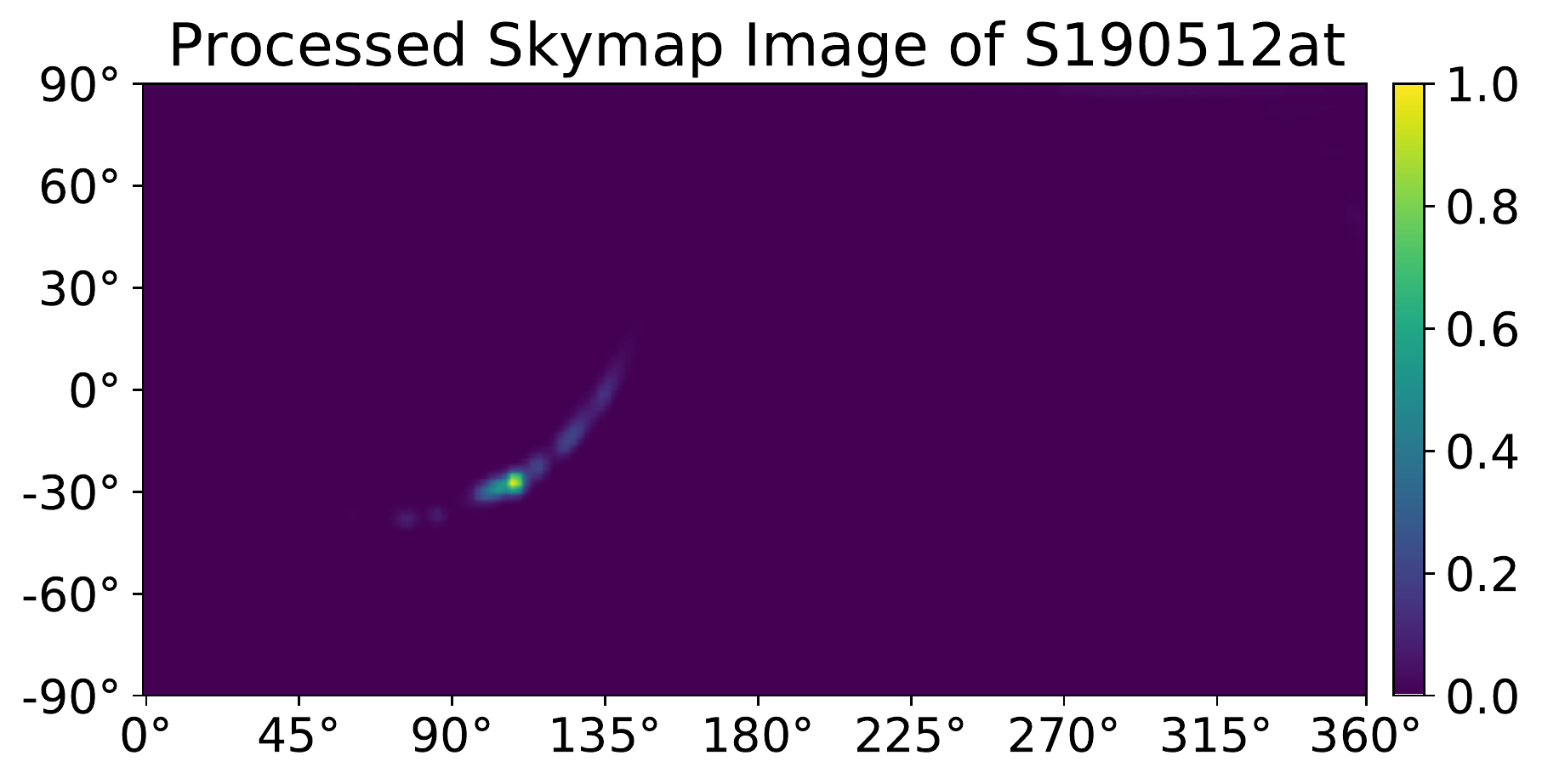}}
    \subfloat[Activation map of the BBH-versus-all classifier on the skymap of S19012at\label{fig: gradcam  b}]
    {\includegraphics[width=0.5\textwidth]{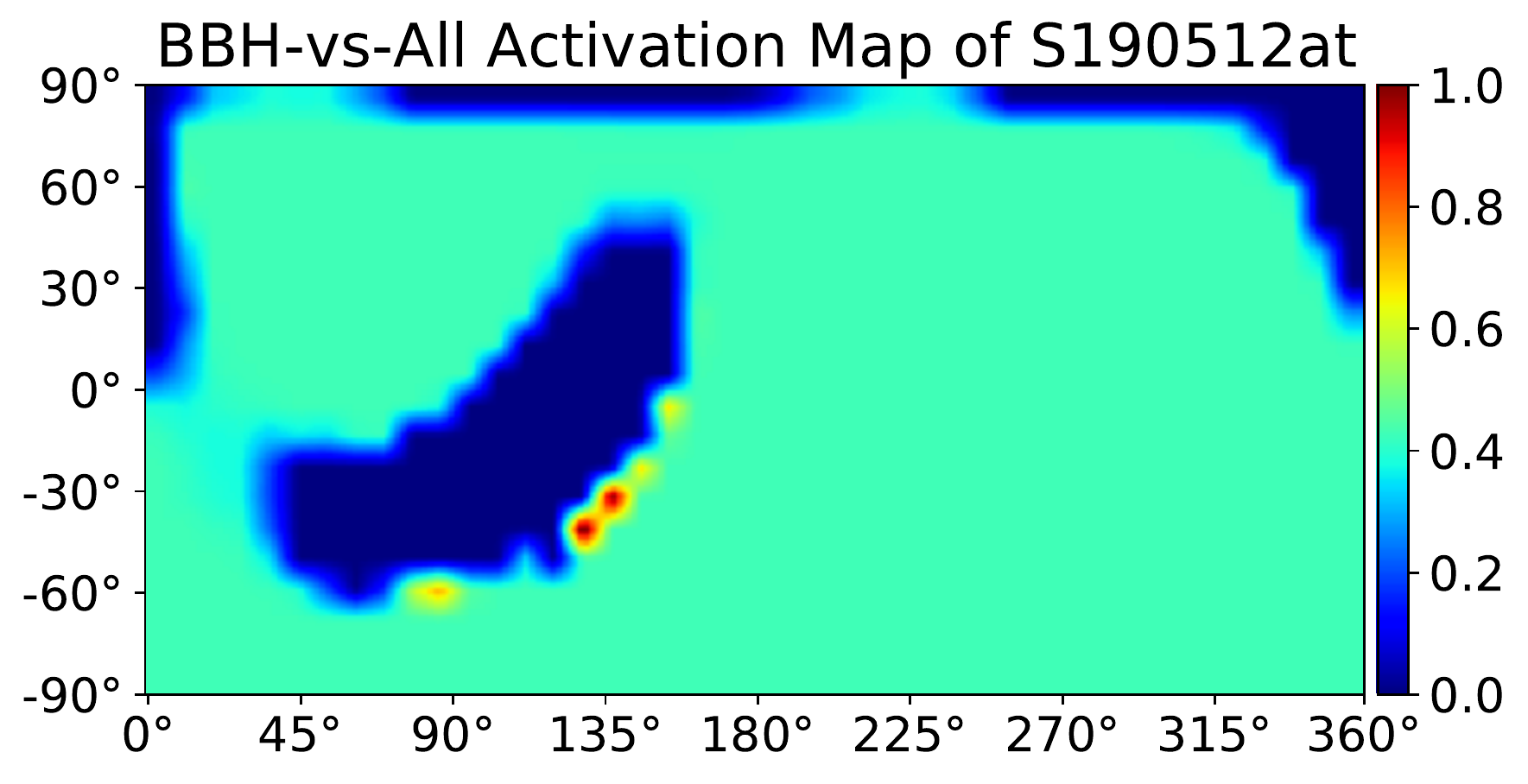}}
    \\
    \subfloat [Processed skymap of the S190923y localization region.\label{fig: gradcam  c}] 
    {\includegraphics[width=0.5\textwidth]{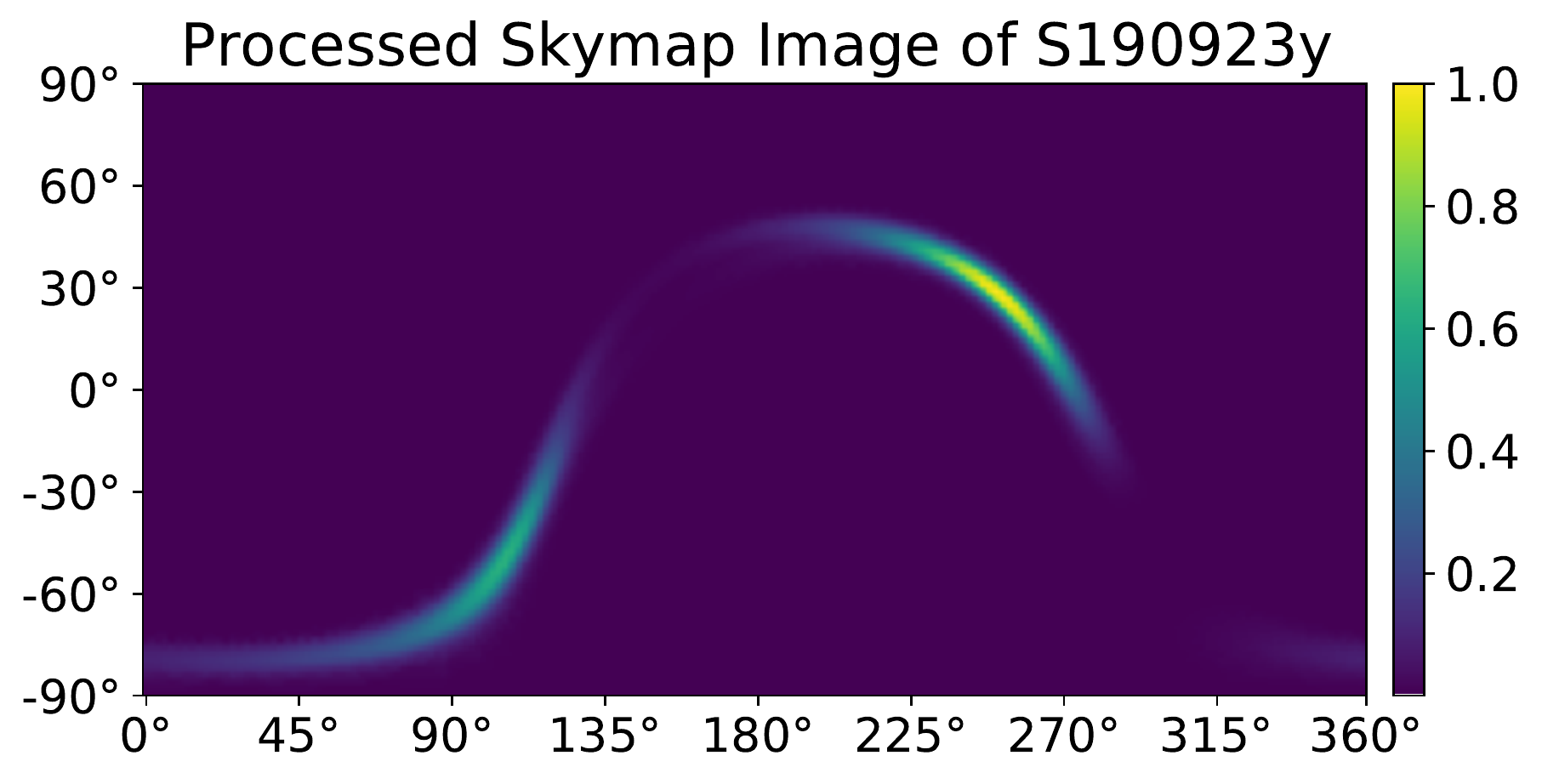}}
    \subfloat [Activation map of the Glitch-versus-all classifier on the skymap of S190923y \label{fig: gradcam  d}]
    {\includegraphics[width=0.5\textwidth]{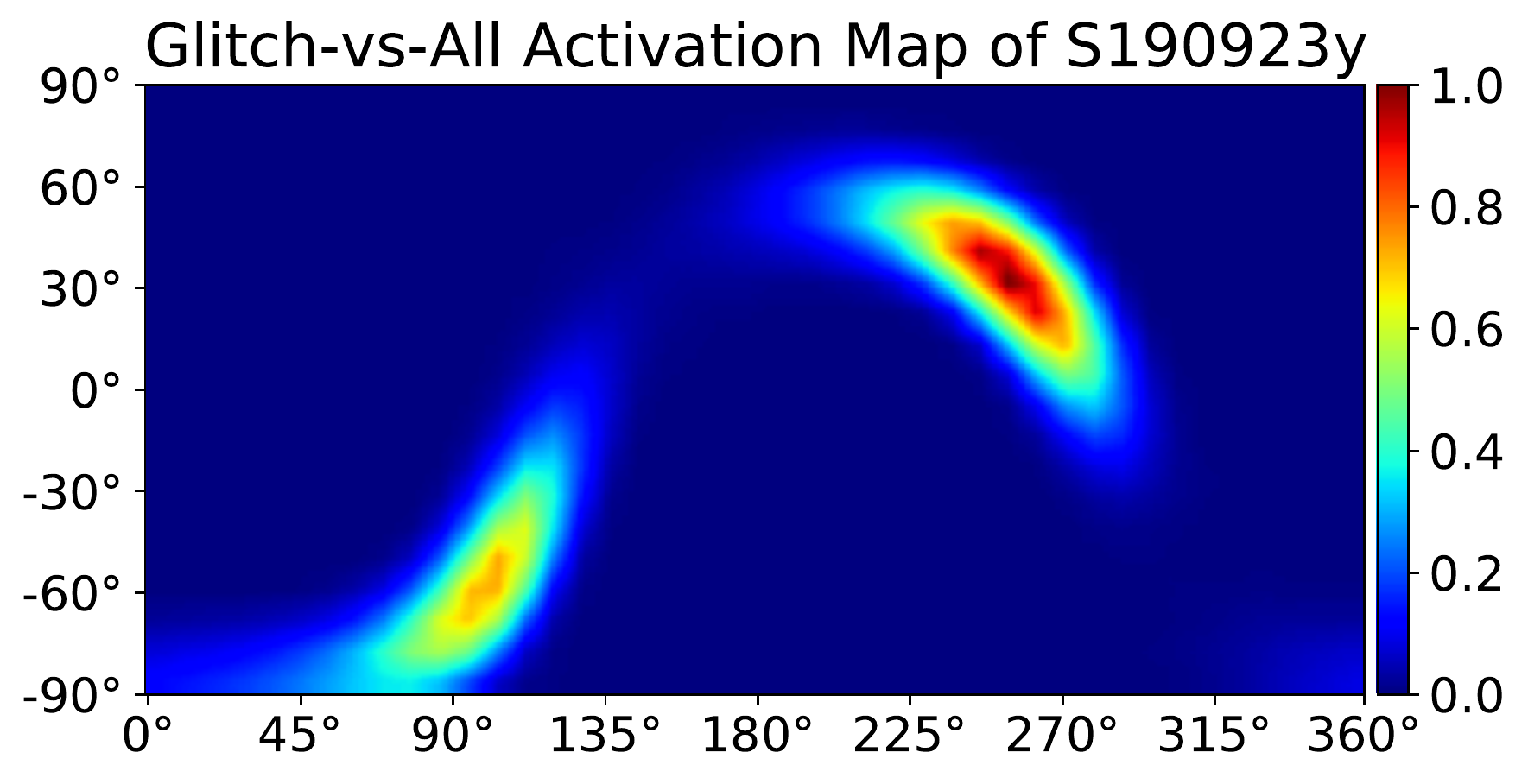}}
    \caption{\textbf{Processed skymaps and Grad-CAM activation maps for S190512at (top) and S190923y (bottom)}. We see that for S190512at, the BBH-versus-all model does not focus on the localization region in the sky, but on the background and the surroundings instead. However, this was a confident and correct prediction, with a BBH-versus-all score of 99.7\%. On the other hand for S190923y, which is a glitch, the glitch model focuses on the localization region but does not classify the event as a glitch. Activation maps are smoothed using a bilinear interpolator.}\label{fig: gradcam weirdos}
\end{figure*}

We also caution that there are notable exceptions to these expectations. For certain sources, the event is confidently and correctly classified, despite the model focusing on the exterior of the high-probability region. We show the example of S190512at in the top panels of Figure \ref{fig: gradcam weirdos}. In contrast, some sources are misclassified despite the relevant models placing all of their focus on the high-probability region of the given skymap. We show the example of S190923y in the bottom panels of Figure \ref{fig: gradcam  weirdos}.

In sum, the activation map alone is not a good indication of the validity of a classification. We leave the exploration of other interpretability techniques for future work.

\section{Conclusion}\label{sec: conclusion}
In this paper, we introduce \texttt{GWSkyNet-Multi}, a multi-class machine learning classifier that enhances the original \texttt{GWSkyNet} by further categorizing GW candidates as BBH mergers, mergers containing one or more NSs, or experimental glitches. \texttt{GWSkyNet-Multi} is created using three one-versus-all classifiers with high accuracy and $F_1$ scores. The glitch-versus-all, NS-versus-all, and  BBH-versus-all classifiers have test set accuracies of 95.1\%, 94.4\%, and 93.7\%, respectively, and all $F_1$ scores are above $0.95$. The classifier also achieves comparable accuracies on the O3a events to \texttt{GWSKyNet} and higher accuracies than GraceDB. These results in addition to the high speed in which the classifiers predict are very encouraging towards the goal of increasing the success rate of EM follow-up of GW events, especially in the upcoming O4 observing run. Furthermore, with the addition of the Kamioka Gravitational Wave Detector (KAGRA), and other GW detectors, OPA skymaps will offer better localizations, which will also greatly increase the likelihood of successful EM follow-up.

\newpage
\section*{Acknowledgements}

The authors acknowledge funding from the New Frontiers in Research Fund Exploration program. N.V. and D.H. acknowledge funding from the Bob Wares Science Innovation Prospectors Fund. D.H.\ acknowledges support from the Canada Research Chairs (CRC) program and the NSERC Discovery Grant program. This research has made use of data, software and/or web tools obtained from the Gravitational Wave Open Science Center (\url{https://www.gw-openscience.org/}), a service of LIGO Laboratory, the LIGO Scientific Collaboration and the Virgo Collaboration. LIGO Laboratory and Advanced LIGO are funded by the United States National Science Foundation (NSF) as well as the Science and Technology Facilities Council (STFC) of the United Kingdom, the Max-Planck-Society (MPS), and the State of Niedersachsen/Germany for support of the construction of Advanced LIGO and construction and operation of the GEO600 detector. Additional support for Advanced LIGO was provided by the Australian Research Council. Virgo is funded, through the European Gravitational Observatory (EGO), by the French Centre National de Recherche Scientifique (CNRS), the Italian Istituto Nazionale di Fisica Nucleare (INFN) and the Dutch Nikhef, with contributions by institutions from Belgium, Germany, Greece, Hungary, Ireland, Japan, Monaco, Poland, Portugal, Spain.

\software{\href{https://docs.astropy.org/en/stable/}{\texttt{astropy}}: \cite{astropy18}; \href{https://lscsoft.docs.ligo.org/ligo.skymap/quickstart/bayestar-injections.html}{\texttt{BAYESTAR}}: \cite{BAYESTAR-code};  \href{https://lscsoft.docs.ligo.org/ligo.skymap/}{\texttt{ligo.skymap}}; \href{https://scikit-learn.org/stable/index.html}{\texttt{scikit-learn}}; \href{https://www.tensorflow.org/}{\texttt{TensorFlow}}}

\clearpage

\appendix{}

\section{Confusion Matrices}\label{app:confusionmatrices}

\begin{figure}[!ht]
    \centering
    \subfloat[Glitch-versus-all confusion matrix.]{\includegraphics[width=0.3\textwidth]{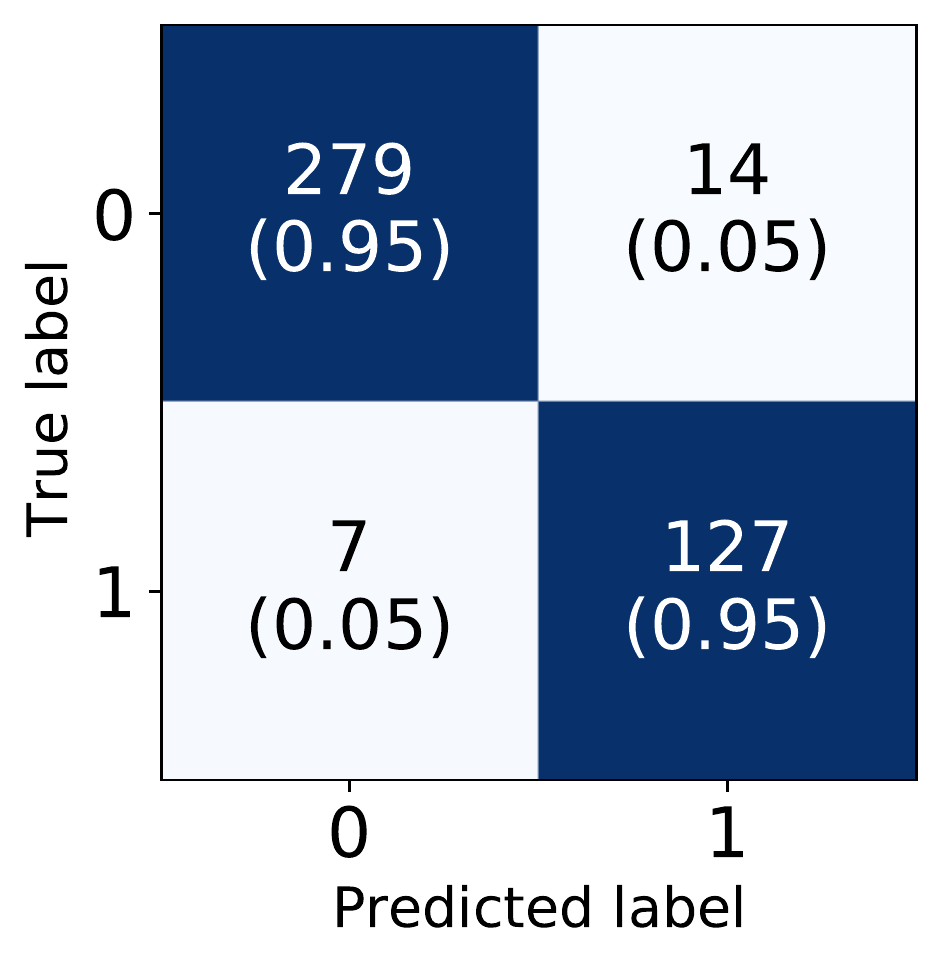}}
    \subfloat[NS-versus-all confusion matrix.]{\includegraphics[width=0.3\textwidth]{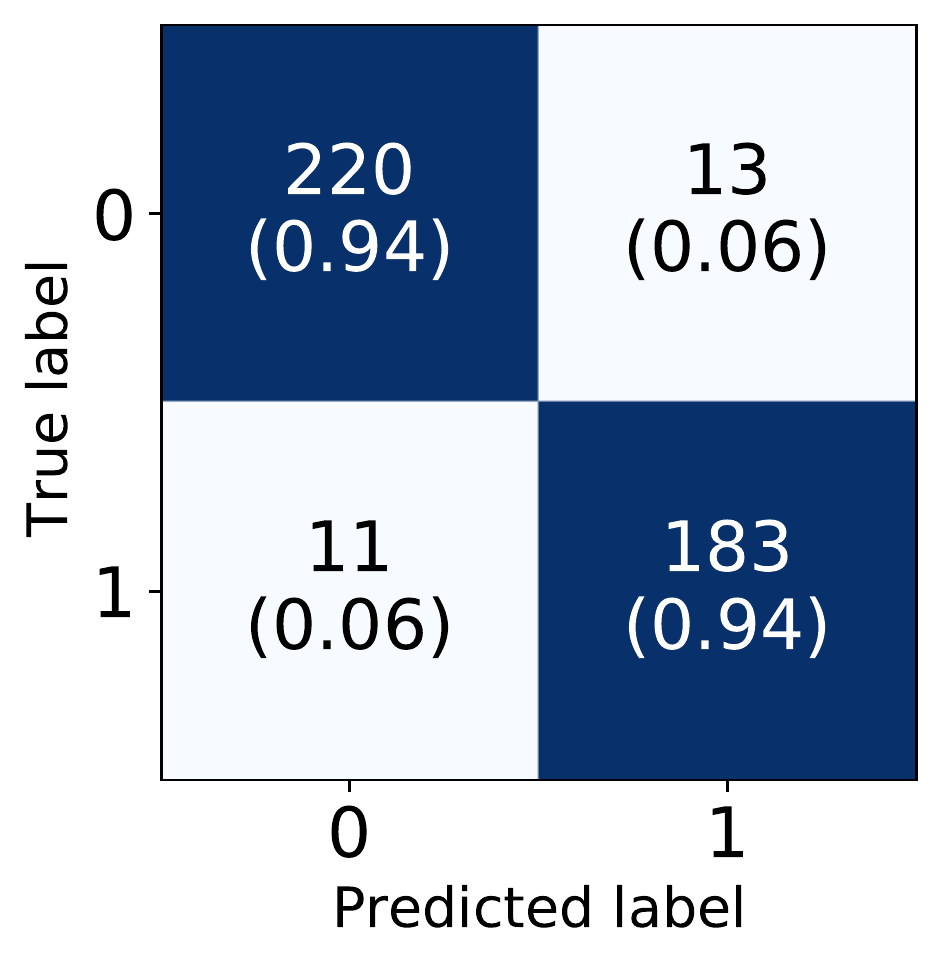} }
    \subfloat[\centering BBH-versus-all confusion matrix.]{{\includegraphics[width=0.3\textwidth]{Glitch_confusion_matrix.pdf} }}
    \caption{Confusion Matrices of each one-versus-all classifier making predictions on the test set.}
    \label{fig: confusion matrices}
\end{figure}

\bibliography{sample631}{}
\bibliographystyle{aasjournal}



\end{document}